\documentclass[12pt,a4paper]{article}
\usepackage[dvips]{graphicx}
\usepackage{times}
\usepackage{rawfonts}
\usepackage{graphics}
\usepackage{natbib}
\usepackage{epsf}
\usepackage{latexsym}
\usepackage{bm}

\usepackage{amsmath}
\usepackage{amssymb}

\usepackage{natbib}
\usepackage{amssymb}
\usepackage{latexsym}
\usepackage{amsmath}
\usepackage{multicol}
\usepackage{graphicx}

\newcommand{\defn}{\begin{quote}{\bf Definition. }}
\newcommand{\edefn}{\end{quote}}
\newcommand{\thm}{\begin{theorem}}
\newcommand{\ethm}{\end{theorem}}

\newcommand{\bp}{{\bm \beta}}
\newcommand{\bmat}[1]{\left [ \begin{array}{#1}}
\newcommand{\emat}{\end{array}\right ]}

\begin{document}

\title{Modelling a response as a function of high frequency count data: the association between physical activity and fat mass}
\author{Nicole H. Augustin,$^{1*}$
Calum Mattocks,$^{2}$
Julian J. Faraway,$^{1}$ Sonja Greven$^{3}$\\ and Andy R. Ness$^{4}$\\
$^1$ Department of Mathematical Sciences, University of Bath, Bath, UK\\
$^2$ Centre for Exercise, Nutrition and Health Sciences,\\
School for Policy Studies,
8 Priory Road,
University of Bristol,\\
Bristol BS8 1TZ\\
$^{3}$ Department of Statistics, Ludwig-Maximilians-University\\ Munich, Ludwigstr. 33,
80539 Munich, Germany\\
$^4$ School of Oral and Dental Science and School of Social\\ and Community Medicine,
Bristol Dental School,\\ Lower Maudlin Street, Bristol, BS1 2LY, UK
}
\maketitle

\begin{abstract}
We present a new statistical modelling approach where the response is a function of high frequency count data. Our application is about investigating the relationship between the health outcome fat mass and physical activity (PA) measured by accelerometer. The accelerometer quantifies the intensity of physical activity as counts per epoch over a given period of time. We use data from the Avon longitudinal study of parents and children (ALSPAC) where accelerometer data is available as a time series of accelerometer counts per minute over seven days for a subset of  children. In order to compare accelerometer profiles between individuals and to reduce the high dimension a functional summary of the profiles is used. We use the histogram as a functional summary due to its simplicity, suitability and ease of interpretation. 
Our model is an extension of generalised regression of scalars on functions or signal regression \citep{MarEil99, RamsSilv05}. It allows also multi-dimensional functional predictors and additive non-linear predictors for metric covariates.
The additive multidimensional functional predictors allow investigating specific questions about whether the effect of PA varies over its intensity, by gender, by time of day or by day of the week. 
The key feature of the model is that it  utilises the full profile of measured PA without requiring cut-points defining intensity levels for light, moderate and vigorous activity. 
We show that the (not necessarily causal) effect of PA is not linear and not constant over the activity intensity. Also, there is little evidence to suggest that the effect of PA intensity varies by gender or whether it happens on weekdays or on weekends.  
\end{abstract}


\section{Introduction}
\label{s:intro}

In many application areas of statistics, high frequency monitoring of counts or intensity occurs. Examples include the monitoring of traffic, heart rate in intensive care, brain activity and financial transactions. 
We are concerned with data on the intensity of physical activity (PA) recorded by an accelerometer. Accelerometers are now widely used for many different applications such as for protecting hard drives from damage in case a laptop is dropped, for monitoring vibrations in industry and in many medical situations. The specific accelerometer used for the application discussed here records the time varying acceleration signal at a specific frequency ranging between 10 to 50 Hz. The signal passes through a filter that band-limits it to a frequency range typical for human motion, excluding activity atypical to human motion such as vibrations from a car. Due to limited storage the signal, measured in units called counts, is summed over a user specified interval called an epoch. The resulting data are hence counts per epoch, ranging between 10 - 60 seconds, over a given period of time, e.g. a number of weeks. Due to data storage becoming smaller and cheaper, newer generation accelerometers are able to store the pass-band filtered raw signal, without summing it over epoch.  This leads to even denser time series of PA measurements. 

In epidemiology accelerometers are now increasingly used to monitor the intensity of PA for investigating its relationship with adiposity, cardiovascular disease, diabetes, depression and other health outcomes.
Depending on the question, different statistical methods are required. For estimating energy expenditure from accelerometer output traditionally regression methods are used. 
More recently machine learning algorithms such as support vector machines \citep{GrueneBroek12,ZhangRow13}, artificial neural networks \citep{StaudPob09,YangZheng2012} and hidden Markov models \citep{PobStau06} have been used for estimating energy expenditure and classifying type of activity. All these methods rely on initial calibration studies which record energy expenditure and accelerometer counts for different type of activities. 

In the epidemiological setting where PA may be a health outcome or the predictor of a health outcome, the high dimensional accelerometer time series are typically summarised into a single summary statistic per individual. Examples for such  statistics are: total activity defined as the average accelerometer counts per minute (cpm), average daily moderate to vigorous PA (MVPA) which is the average minutes per day spent at moderate or vigorous activity and average sedentary behaviour is defined as the average minutes per day spent in sedentary activity \citep{Rid09,Mitch09,Mitch11}.  For each of these summaries cut-points of counts per minute are used for different levels of energy expenditure relating to light, moderate and vigorous activity. These are based on estimates of energy expenditure described above, either using published models from the literature or using models based on calibration studies performed on a subset of the study population. There is some debate whether the energy expenditure changes with age, in particular for children, hence the cut-points may change with age \citep{Reilly05}. Using only scalar summaries ignores the pattern of PA, meaning  the distribution of the activity counts per minute and the frequency spectrum of activity counts. 

Here we present a model with a scalar response regressed on additive multidimensional functional predictors for exploring the relationship between PA and  health outcomes.  The major advantage of the model is that it allows us to take the pattern of PA into account making it cut-point independent. We use the modelling approach to investigate the following questions: 
\begin{enumerate}
 \item Is the effect of PA linear over the activity intensity range, i.e. is it sufficient to use average counts per minute as a predictor? or does the effect of PA vary over the intensity range and how?
 \item Does the effect of PA intensity vary by gender? This would be plausible as there are differences in the metabolism.
 \item Does the effect of PA intensity vary by time of day?  
 \item Does the effect of PA intensity vary depending on whether it happens on a weekday or on  the week-end? Evidence for different effects could be explained by the fact that weekend and weekday are associated with different activity types.
\end{enumerate}

We develop our approach for investigating the above questions using data from the Avon longitudinal study of parents and children (ALSPAC) where accelerometer data is available as a time series of accelerometer counts per minute over seven days for a subset of  children. Our interest lies in fat mass as a health outcome. At ages of approximately 12, 14 and 16 the children were attending research clinics where they were asked to wear an accelerometer for 7 days. Fat mass was also assessed using a DXA scanner. 
Regarding the association between fat mass and PA the data have been extensively analysed, by considering summary statistics such as MVPA and total PA as predictors for the health outcome fat mass and vice versa. The results indicate that MVPA is associated with a lower risk of obesity in children as they go through adolescence and more so in boys than girls \citep{Mattocks11}. 

The remainder of this article is organised into four sections. In section~\ref{secdat} we introduce the data, the processing protocol and present summary statistics.  In section~\ref{secmeth} we present the regression of a scalar on multi-dimensional additive regression functions model.  Then we present results (section \ref{sec:results}), followed by a discussion (section \ref{sec:discuss}).

\section{The data}
\label{secdat}
ALSPAC is a birth cohort study, see \cite{BoyGolMac12} and \cite{FrasMacDTill2013} for a detailed description. In this study, all pregnant women in the former Avon Health Area who had an expected delivery date between April 1, 1991 and December 31, 1992, were asked to participate in the study. The Avon Health Area is situated around Bristol in the UK. A total of 14, 541 pregnant women were enrolled,  and this resulted in 14,062 live births.  Detailed data have been collected by self-completed questionnaires from pregnancy onward. All children of the study have been invited to regular research clinics from the age of 7. At the ages of 11, 13 and 15 the children attending these clinics were asked to wear an MTI Actigraph AM7164 2.2 Accelerometer for 7 days. As the mean ages of children were actually closer to 12, 14 and 16 years we will refer to the children as 12, 14 and 16 year olds.   Children wore the accelerometer during waking hours, except for showering, bathing, swimming and any water sports. At age 14 children were wearing either the AM7164 or the newer GT1M Actigraph. 
The Actigraph AM7164 2.2 is a uni-axial piezoelectric accelerometer which records acceleration in the vertical direction. The acceleration is measured 10 times per second (10 Hz) and the resulting signal is filtered to band-limit it to the frequency range 
excluding activity untypical to human motion such as vibrations from a car. This accelerometer requires regular unit calibration. Due to limited storage the signal is converted into counts and summed over a user specified interval called an epoch, e.g. 1 minute.  The Actigraph GT1M accelerometer also used at age 14 is a more advanced model. Using a solid state monolithic accelerometer it samples the signal at 30 Hz and band-limits the signal using a digital filter. 
Unfortunately data on which type of accelerometer was worn by the children at age 14 was not available for this analysis.

Fat mass was derived using a Lunar Prodigy DXA scanner (GE Medical Systems Lunar, Madison, WI, USA). In addition the children's height was measured.
 See \cite{MatNesLea08} for more details on the measurement protocol.

Please note that the study website contains details of all the data that is available through a fully searchable data dictionary, see\\ 
{\tt http://www.bris.ac.uk/alspac/researchers/data-access/data-dictionary/}.
\subsection*{Processing of activity profiles}

The accelerometers were set to an epoch time of 1 minute and hence for each child there was a time series of minute by minute accelerometer measurements (counts per minute) over seven days available at the ages 12, 14 and 16.
For preprocessing of the accelerometer data we followed \cite{MatNesLea08}. Any sequence with more than 10 zeros was replaced by missing values, since these periods were regarded as periods were the monitor was not worn; days with a mean count less than 150 or a mean count of three standard deviations above the overall mean (prior to exclusions) were invalid; days were only included if the monitor was worn at least 600mins  (10h); weekly profiles were invalid if less than 3 valid days were observed. 

Previously, protocols for processing accelerometer data have varied, although there is a move towards standardisation. For instance to treat a block of 10 zeros as missing, i.e. non-wear time, leads to a lower overall wear time and may lead to a lower sample size compared to treating blocks of 60 zeros or more as missing. Hence in other studies often blocks of 60 zeros or more are treated as non-wear times. These non-wear times could be due to above average intensity activities where the accelerometer was not worn to prevent damage including swimming and contact sports. As in the ALSPAC study participants were asked to record activities during periods when the accelerometer was not worn, the information from these self-reported activities was incorporated into estimates of MVPA by \citep{GriffLear12} to investigate possible effects of measurement error. Incorporating the self-reported activity increased the time spent in moderate to vigorous activity at 12 years from 20 to 25 minutes.  Results showed that the effect of PA was weakened when the self-reported activities were incorporated into the estimates of MVPA. Possible reasons for this counter-intuitive result are that self-reported PA may overestimate levels of activity. For this reason, we use only PA measured by accelerometer here, being aware of possible biases caused by missingness due to non-wear time.

Activity counts greater than 15000 cpm were set to missing, as it is unclear what activity would result in such high counts, and misrecording of activity was most likely. This is in line with two independent validation studies using  MTI Actigraph AM7164 \citep{GriffMatt12, Brage03} where 
intensity for running or jogging was recorded at a maximum value of around 12500 cpm.

\subsection{Summary statistics}
Here we consider fat mass and activity profiles at all three ages 12, 14 and 16. Not all of the children who had their fat mass measured by scanner also had their PA measured and vice versa. Following the processing protocol of the accelerometer data this yields 4161 children with both measurements at age 12 and 2919 children at age 14 and 1488 children at age 16.  
Below we use the data resulting from the block of 10 zero rule for non-weartime.
 
\begin{figure}[!h]\centering
\includegraphics[scale=0.4,angle=-90]{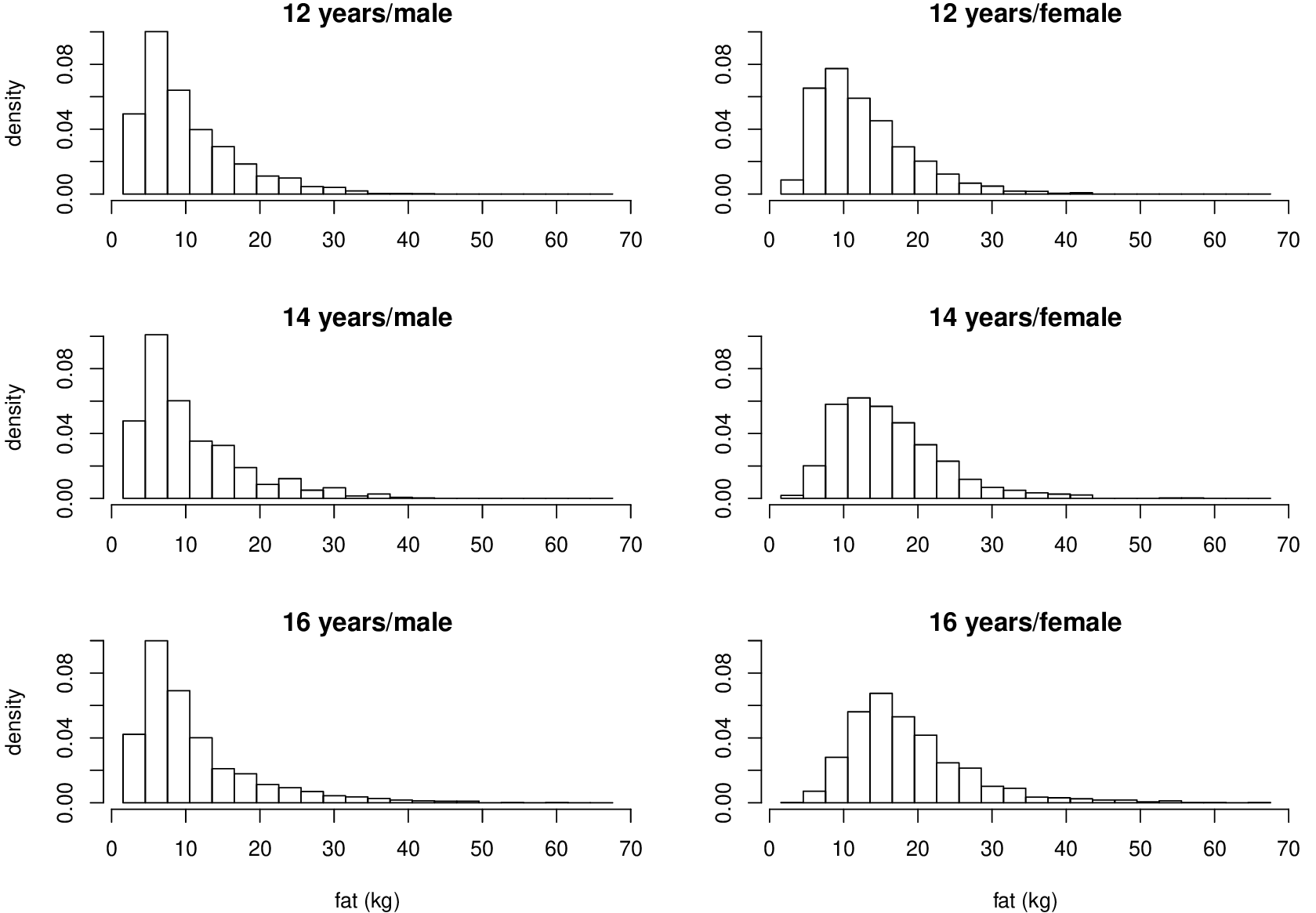}\\
\vspace{1cm}
\includegraphics[scale=0.4,angle=-90]{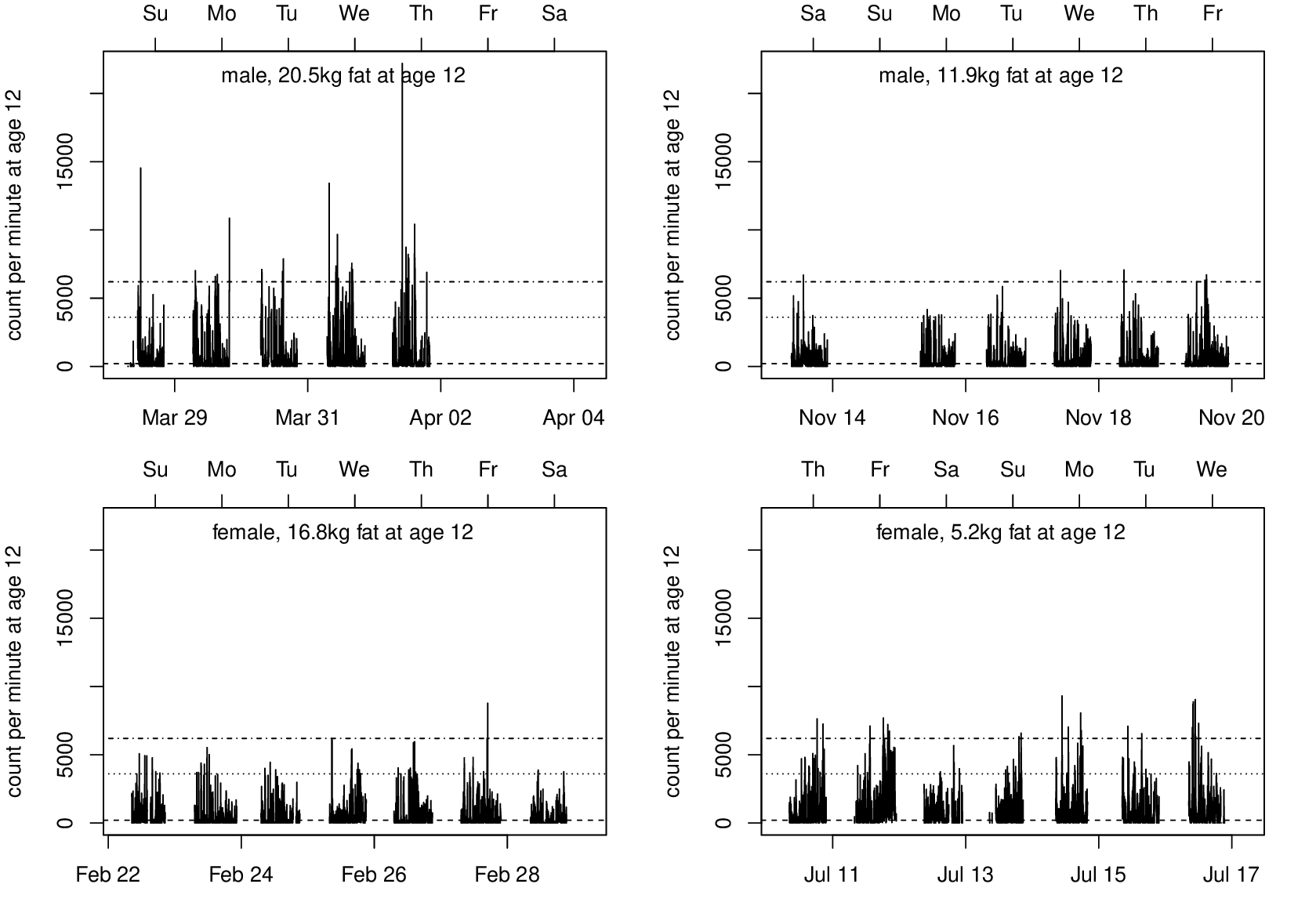}
\caption{Top three rows: Histogram of fat mass (kg) measured by scanner at ages 12, 14 and 16 for boys and girls. Bottom two rows: Sample of 4 PA profiles at age 12. The horizontal lines denote from bottom to top the cut points for light, moderate and vigorous PA at 200, 3600, 6200 counts per minute derived in the calibration study \citep{MatLeaNes07}.}
\label{fig:fat}
\end{figure}

The distribution of the outcome variable fat mass (kg) is right-skewed (Figure~\ref{fig:fat}). Girls have a higher fat mass than boys and this difference becomes more pronounced as they go through adolescence. 

Figure~\ref{fig:fat} shows 4 samples of the accelerometer profiles at age 12 in the processed form, that is blocks of 10 zeros and invalid days have been set to missing. 
It is apparent that  individuals cannot be compared directly with these time series of length 10080 and some dimension reduction is required. A suitable dimension reduction is the histogram which summarises the distribution of the activity counts per minute.
%
\begin{figure}[!h]\centering
\includegraphics[scale=0.5,angle=-90]{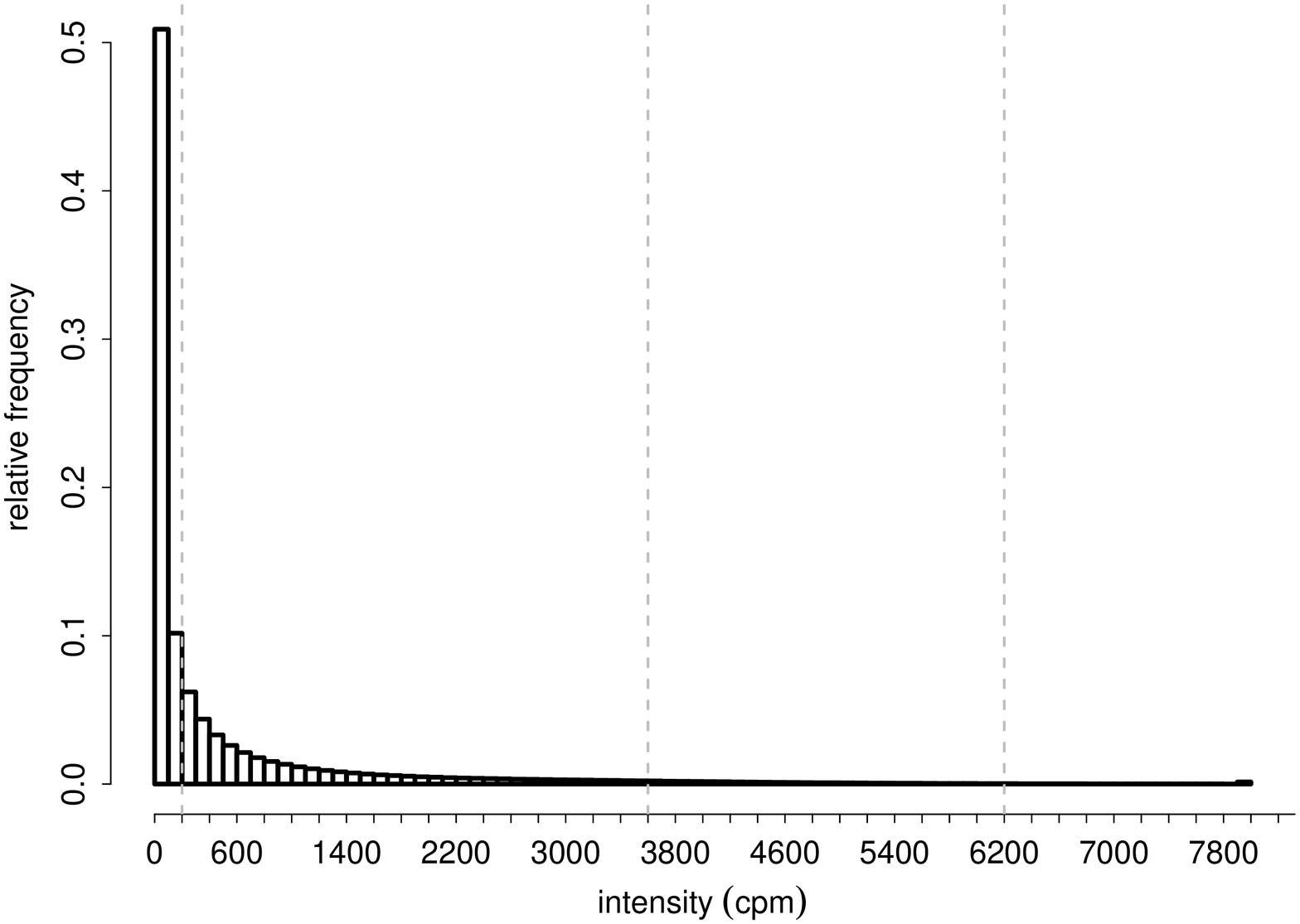}
\caption{The mean histogram (relative frequency) with 80 bins over all individuals and ages. The cut-points for light, moderate and vigorous PA currently used in ALSPAC \citep{MatLeaNes07} are at 200, 3600, 6200 counts per minute. 
}
\label{fig:sumfun}
\end{figure}
Figure~\ref{fig:sumfun} shows the mean histogram over all individuals and ages. 80 bins of equal width were used with the exception of the last bin which contains all counts in the range [8000,15000). A plot of the difference between the mean histogram of each age by gender combination (not shown) gives the following insight. At the same age on average boys are more active than girls. The biggest differences in intensity distribution are seen between ages, at age 12 children spent the smallest amount of time in sedentary activity (very low to zero cpm) and spent more time in moderate to vigorous activity. With increasing age boys and girls are on average less active.

\begin{figure}[!h]\centering
\includegraphics[scale=0.3,angle=-90]{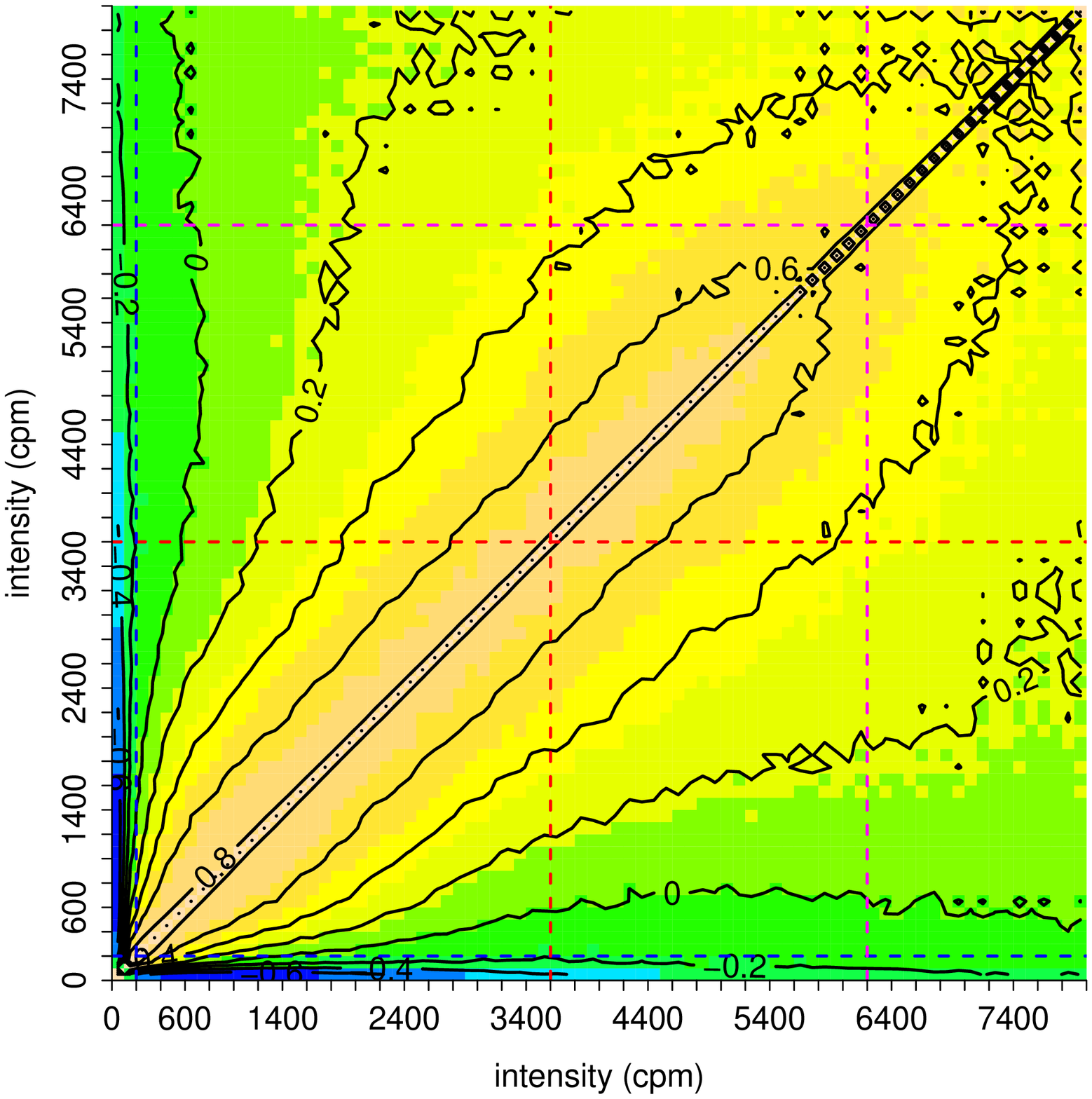}\includegraphics[scale=0.3,angle=-90]{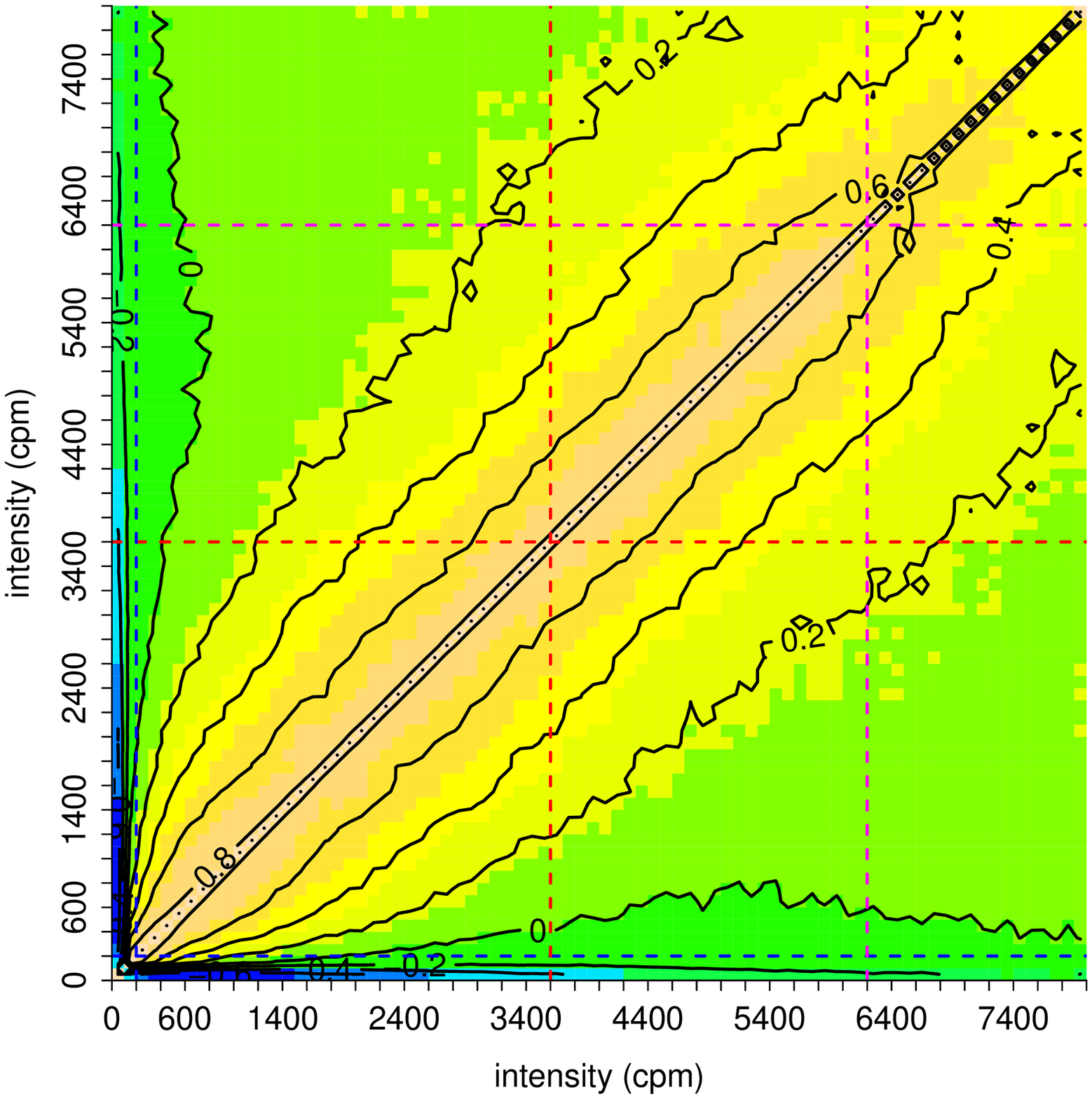}\\
\includegraphics[scale=0.3,angle=-90]{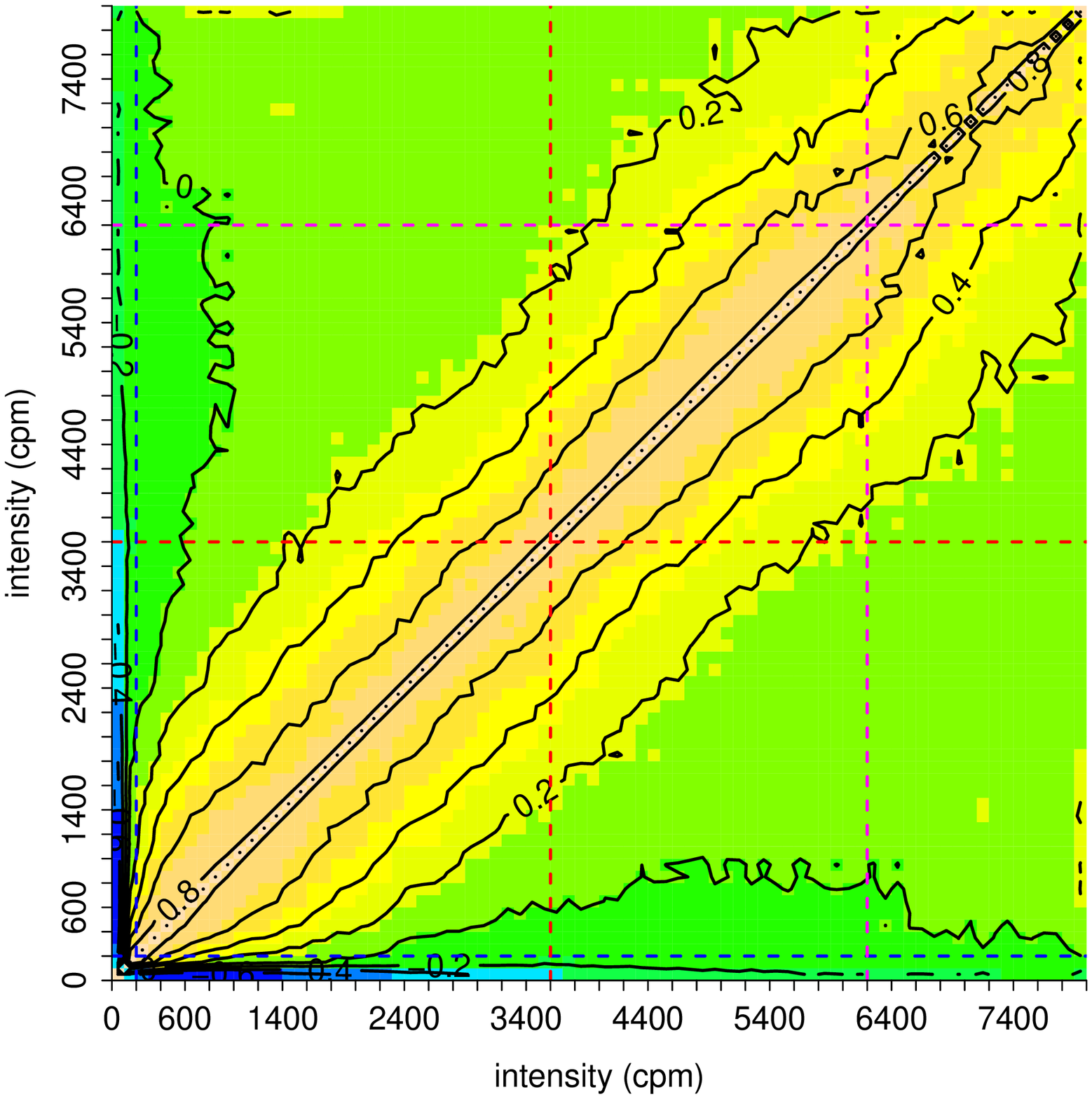}
\caption{Estimated correlation between the density of the histogram with 80 bins for individuals at age 12, 14 and 16. The dashed lines denote PA at 200, 3600, 6200 counts per minute. The labeled isolines denote the correlation. The color scheme is from dark blue (correlation $<$ -0.6), via green, yellow, orange (correlation $>$ 0.6) to white (correlation = 1).}
\label{fig:cov}
\end{figure}

Figure~\ref{fig:cov} shows the correlation between the density of the histogram bins for individuals at age 12, 14 and 16. Besides very strong positive correlation between density values of neighbouring bins, there is negative correlation  up to -0.6 between sedentary behavior (below the first cut-point line) and light activity (above the first cut-point line). 
For all three ages the correlation between sedentary and moderate to vigorous activity is very small. This implies that if 
individuals spent little time in sedentary activity they spent more time in light activity and vice versa, if individuals spent a lot of time in sedentary activity they spend less time in light activity. 
%
\section{Models}
\label{secmeth}

We introduce the approach by considering the following log-linear model for fat mass as a function of the accelerometer profile
\begin{equation*}  \log(y_{i}) = x_i^T \bp  + \sum_j \gamma_j z_{i}(p_j)+ \epsilon_{i}.
\label{linmod}
\end{equation*}
The response $y_{i}$ is total fat mass for individual $i$ at a given age, $z_{i}(p_j)$ is the  relative frequency of the histogram with some given number of mid-points $p_j$ and  derived from the vector of 10080 entries.   The error $\epsilon_{i} \sim N(0, \sigma^2)$ and the vector $x_i$ is a row of the model matrix containing scalar variables and factors ${\tt sex, height, height}^2, {\tt weartime}$ 
and whether the mother is obese (${\tt m.obese}$). The variable ${\tt weartime}$ is included because it is a measure of compliance which could be a proxy for hidden confounders. The problem with this model is that, besides being unidentifiable due to $\sum z_i(p_j)=1$, the coefficients $\gamma_j$ for the histogram values will be highly correlated and their interpretation will hence be difficult. The model treats the relative frequency for each of the histogram mid-points as a separate explanatory variable and this is not appropriate. It makes more sense to exploit the fact that the histogram is a function and constrain the estimate of $\gamma_j$ to be similar to the $\gamma_l$ at neighbouring mid-points $p_l$. That is the histogram enters the model as a function.  We let the $\gamma_j$ vary smoothly, where $\gamma_j = f(p_j)$, an unknown `coefficient' function to be estimated and $f(p)$ can be parameterised using any penalised regression spline basis. Then  
\begin{equation*}  \log(y_{i}) = x_i^T \bp  + \sum_j f(p_j) z_{i}(p_j)+  \epsilon_{i}.
 \label{genreg}
\end{equation*}
Here $f(p)$ is represented using an adaptive smooth with a P-spline basis \citep{EilMar96},  based on a B-spline basis function and discrete penalties on the basis coefficients. The model is identifiable because the smooth function $f(p)$ is centred, i.e. the constraint $\sum_j f(p_j)=0 $ is imposed. The smooth is adaptive by letting the terms in the penalty have different weights depending on the histogram mid-point $j$ and then multiplying the smoothness parameters by these weights.
Models like above are also called  generalised regression of scalars on functions or signal regressions see for example \cite{MarEil99} or \cite{RamsSilv05}.

\subsection*{Model selection}
We extend the above model by allowing also multi-dimensional functional predictors and additive non-linear predictors for metric covariates. This is illustrated by the following models, each of which will be used to answer one of the four questions posed in the introduction. 
\begin{equation} \mbox{Question 1~~~~~~~~~~~~~~~~~~~~~~~~~~~~~~~~} \log(y_{i}) = x_i^T \bp + f_2({\tt height}_i)+   \sum_j f(p_j) z_{i}(p_j) + \epsilon_{i}
 \label{modbasic}
\end{equation}
where $f_2$ is represented using a thin plate regression spline and the vector $x_i$ contains ${\tt sex, weartime}$ and ${\tt m.obese}$.  Note the special case of model~\ref{modbasic}:
\begin{equation}  \log(y_{i}) = x_i^T \bp + f_2({\tt height}_i)+  \gamma  \sum_j p_j z_{i}(p_j) + \epsilon_{i},
 \label{modlin}
\end{equation}
with a linear effect of the mean counts per minute $\sum_j p_j z_{i}(p_j)$ estimated from the histogram. 

Further we can investigate in a later step whether the effect of PA intensity varies by gender by introducing different smooth functions for males and females: 
\begin{equation} \mbox{Question 2~~~~~~~~~~~~~~~~~~~~~~~~~~~~~~}  \log(y_{i}) = x_i^T \bp + f_2({\tt height}_i)+  \sum_j f_{s(i)}(p_j) z_{i}(p_j) + \epsilon_{i}
 \label{modsex}
\end{equation}
where $s(i)$ = 1  for male and $s(i)=2$ for female. For investigating whether the effect of PA intensity varies with time of day we introduce a two dimensional histogram over intensity and time of day:
\begin{equation} \mbox{Question 3~~~~~~~~~~~~~~~} \log(y_{i}) = x_i^T \bp + f_2({\tt height}_i)+ \sum_j \sum_m f(p_j,t_m) z_{i}(p_j,t_m) l_m + \epsilon_{i}
 \label{mod2D}
\end{equation}
where $z_{i}(p_j,t_m)$ is a two dimensional histogram of PA intensity with bins of equivalent width as in \ref{modbasic} and hour of the day with bin widths $l_m$. The smooth function  $f(p_j,t_m)$ is now a two-dimensional smooth, expressed by a tensor product of two cubic regression spline bases, where we apply separate penalties for time and intensity.
And lastly, we investigate whether the effect of PA intensity varies between weekday and weekend.
\begin{equation}\mbox{Question 4~~~}  \log(y_{i}) = x_i^T \bp + f_2({\tt height}_i) + \sum_j f_{we}(p_j) z_{i}^{we}(p_j)+\sum_j f_{wd}(p_j)  z_{i}^{wd}(p_j)+ \epsilon_{i}
 \label{modWE}
\end{equation}
where $f_{we}$ is the smooth function over the histogram for the weekend and $f_{wd}$ for weekdays, with separate histograms for week-end $z_{i}^{we}(p_j)$ and week day $z_{i}^{wd}(p_j)$.

We use the root mean square prediction error (RMSPE) estimated by validation as a selection criterion. 
In addition we consider the Akaike information criterion (AIC), Bayesian information criterion (BIC) and adjusted R$^2$.
Comparing these criteria between model~\ref{modbasic} and model~\ref{modlin} will show how much can be gained by allowing the effect of PA intensity to be non-linear. We also test formally whether the effect for PA is non-linear using a significance test. Replacing the adaptive smooth basis with a thin plate regression spline basis makes this task relatively easy. The linear part of the regression spline basis, that is the completely smooth basis function with zero penalty, is separate from the rough penalised basis functions. Hence the linear basis part can be included as a separate effect in a model which also includes the rough penalised basis function. In this nested model we test whether the non-linear part of the basis is zero using an approximate likelihood ratio test~\citep{Wood13}.

\subsection*{Model interpretation}
Regarding the interpretation of model~\ref{modbasic} and its variant models \ref{modsex} - \ref{modWE} it helps to consider the model $\alpha+ \sum_j f(p_j) z(p_j)$ for the scenario that PA is equally distributed along intensity with $z_i(p_j)= c$. Then the intercept $\alpha$ is the expected log fat mass because of the centred smooth constraint $\sum_j f(p_j)c = 0$. The functional coefficient $f(p_j)$ measures the deviation from this expected value according to the actual PA distribution.  

If we insert $z_i(p_1) = 1 - \sum_{j>1} z_i(p_j)$ into the model above, then 
$\alpha +  f(p_1)z_i(p_1) + \sum_{j>1} f(p_j) z_i(p_j)  = \alpha + f(p_1) + \sum_{j>1}  {\left( f(p_j) - f(p_1) \right) z_i(p_j)} = \alpha^{'}+ \sum_{j>1}  f(p_j)^{'} z_i(p_j) $. This shows that if we remove the first histogram bin, the intercept is $\alpha^{'} = \alpha + f(p_1)$, the expected value of log fat mass for sedentary activity, as defined by the breaks of the first histogram. The interpretation of the  functional coefficient is also more convenient: $f(p_j)^{'} = \left( f(p_j) - f(p_1) \right)$  is the effect of activity compared to sedentary activity. Hence in the following analysis we fit all models with the first histogram bin removed instead of imposing the centred smooth constraint to achieve this convenient interpretability.

\subsection{Parameter estimation}
Parameter estimation of the above regression with additive multidimensional functional predictors works as for any other penalised GLM.  Here we use the computational set up of \cite{Wood11} implemented in the {\tt gam()} function of the  {\tt R package mgcv} \citep{Rman11} to estimate the model. The parameters are estimated using a nested iteration scheme \citep{Wood11} where the  outer iteration is approximate restricted maximum likelihood (REML) estimation of smoothness parameters, minimising a Laplace approximation of REML where the parameter vector, containing coefficients of linear terms and of basis functions for the smooth terms, is integrated out. 
The inner iteration is a penalised iterative re-weighted least squares (PIRLS) algorithm to find all other parameters, i.e. the coefficients of basis functions, and coefficients of linear terms.
For prediction we use Bayesian credible intervals
by sampling from the posterior of all model parameters to obtain a sample from the predictive distribution see also \citep{Wahba83, Silv85, Wood06}. For this, the model is represented as a Bayesian model, where the smooth terms are a mixture of fixed and random effects. 
This approach recognises that imposing a particular penalty effectively imposes some prior beliefs about the likely
characteristics of the correct model. That is, the model structure allows considerably more
flexibility than is believed to be really likely, and the choice is made to penalise models that
are in some sense too wiggly. From the sampled parameter vectors we obtain samples of fitted values, i.e. the predictive distribution, by multiplying the model or prediction matrix with each sample parameter vector. See also \cite{AugTrWo13} for a detailed description. 


\section{Results}
\label{sec:results}
We fit model~\ref{modbasic} to \ref{modWE} for 6 different age combinations (a - f) of the variables (Table~\ref{Tab:models}).
\begin{table}[!h]
\begin{center}
\caption{Combinations of variables regarding age used in models.}
\label{Tab:models}
\small
\begin{tabular}{|l|r|r|r|r|r|r|r|r|r|r|r|r|} \hline
\multicolumn{1}{|c|}{variable}&\multicolumn{3}{|c|}{fat mass}&\multicolumn{3}{c|}{height}&\multicolumn{3}{c|}{weartime}&\multicolumn{3}{c|}{PA}\\ \hline
at age &12&14&16&12&14&16&12&14&16&12&14&16\\ \hline
(a)&x&&&x&&&x&&&x&&\\ \hline
(b)&&x&&&x&&&x&&x&&\\ \hline
(c)&&&x&&&x&&&x&x&&\\ \hline
(d)&&x&&&x&&&x&&&x&\\ \hline
(e)&&&x&&&x&&&x&&x&\\ \hline
(f)&&&x&&&x&&&x&&&x\\ \hline
\end{tabular}
\end{center}
\end{table}
We compare these models with the different criteria in Table~\ref{Tab:results}.  The models are fitted to approximately 3/4 of data and the root mean squared prediction error (RMSPE) is then estimated on the remaining 1/4 of data. All the results are based on using 80 bins for the individual histograms of the accelerometer counts. The results show that for most age combinations model~\ref{modbasic} is best, and there is little evidence that any of the more complex models~\ref{modsex} - \ref{modWE} provides a better fit. In (b) models~\ref{modsex} and \ref{mod2D} yield slightly lower RMSPEs than model~\ref{modbasic} and in (f), the AIC, BIC, and RMSPE select different best models, with model \ref{modWE} yielding the lowest RMSPE.

The results also show that the effect of PA intensity is not linear since model~\ref{modbasic} is always better than model~\ref{modlin} which only uses mean count per minute.  Hypothesis testing on whether the effect of PA is non-linear compared to being linear confirms this. For these tests we used a thin-plate regression spline basis for $f(p_j)$ rather than an adaptive smooth. This  allowed to separate out the linear part of the basis and hence include separate effects for linear and non-linear parts of the smooth.

\begin{figure}[!h]\centering
\includegraphics[scale=0.57,angle=-90]{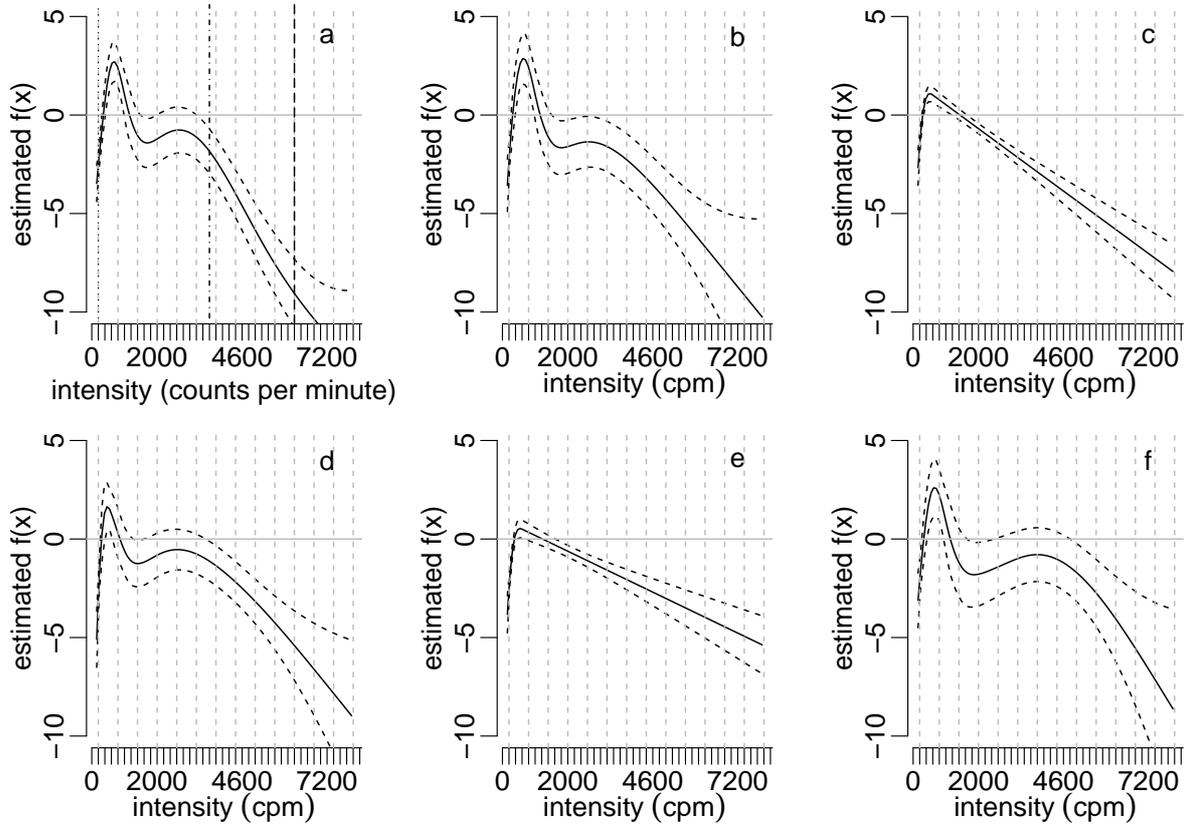}
\caption{The estimated functions $f(p_j)$ for all 6 models. (a) fat mass, height and weartime and PA at age 12; (b) fat mass, height and weartime at age 14 and PA at age 12; (c) fat mass, height and weartime at age 16 and PA at age 12; (d) fat mass, height, weartime and PA at age 14; (e) fat mass, height and weartime  at age 16 and PA at age 14; (f) fat mass, height, weartime and PA at age 16. The gray dashed vertical lines are at 200, 800, 1400, .... In (a) the cut-points for light, moderate and vigorous PA from \citep{MatLeaNes07} estimated at 200, 3600, 6200 counts per minute are also shown.}
\label{fig:est}
\end{figure}

\begin{table}[!h]
\begin{center}
\caption{Model selection results for models using a histogram with 80 bins. For the age combinations (a) - (f) described in Table~\ref{Tab:models} models are compared to model 'base' ($\log(y_{i}) = x_i^T \bp + f_2(height_i) + \epsilon_{i}$) which does not include a term for PA. edf are the effective degrees of freedom of the model, R.sq is the adjusted R$^2$. For AIC and BIC the difference to the 'base' model is given. Parameters, adj. R$^2$, AIC and BIC are estimated from the training data with sample size N$_T$. The root mean squared prediction error (RMSPE) is in kg fat mass and estimated on the validation data.}
\label{Tab:results}
\small
\begin{tabular}{|l|l|r|r|r|r|r|r|} \hline
\multicolumn{1}{|c|}{age combination}&\multicolumn{1}{|c|}{model}&\multicolumn{1}{|c|}{N$_T$}&\multicolumn{1}{c|}{edf}&\multicolumn{1}{c|}{adj. R$^2$}&\multicolumn{1}{c|}{$\Delta$ AIC}&\multicolumn{1}{c|}{$\Delta$ BIC}&\multicolumn{1}{c|}{RMSPE (kg)}\\ \hline
(a)&base&2918&~~~8&~~25.080&~~~0&~~~0&~~~5.895\\ 
&+ hist (1)&2918&~~13&~~31.270&~301.900&~162.200&~~~5.674\\ 
&+ cpm&2918&~~~9&~~28.060&~117.100&~110.900&~~~5.841\\ 
&+ hist by gender&2918&~~16&~~31.120&~292.900&~136.800&~~~5.682\\ 
&+ 2Dhist&2918&~~28&~~31.810&~255&~138.100&~~~5.706\\ 
&+ hist by WE&2918&~~14&~~29.760&~237.200&~~90.100&~~~5.743\\ \hline
(b)&base&1736&~~~6&~~23.330&~~~0&~~~0&~~~6.625\\ 
&+ hist (1)&1736&~~12&~~28.050&~145.900&~~36.250&~~~6.442\\ 
&+ cpm&1736&~~~7&~~25.250&~~42.890&~~37.280&~~~6.587\\ 
&+ hist by gender&1736&~~14&~~27.790&~136.900&~~12.430&~~~6.416\\ 
&+ 2Dhist&1736&~~18&~~27.770&~~91.720&~~27.430&~~~6.417\\ 
&+ hist by WE&1736&~~12&~~26.300&~104.400&~~-4.806&~~~6.468\\ \hline
(c)&base&1630&~~~7&~~24.180&~~~0&~~~0&~~~7.055\\ 
&+ hist (1)&1630&~~10&~~27.900&~121.500&~~19.730&~~~6.956\\ 
&+ cpm&1630&~~~8&~~25.980&~~37.940&~~32.110&~~~7.062\\ 
&+ hist by gender&1630&~~13&~~27.950&~119.800&~~~2.379&~~~6.959\\ 
&+ 2Dhist&1630&~~20&~~27.190&~~53.370&~-15.200&~~~6.992\\ 
&+ hist by WE&1630&~~11&~~26.230&~~83.290&~-23.430&~~~7.047\\ \hline
(d)&base&2156&~~~5&~~32.730&~~~0&~~~0&~~~7.564\\ 
&+ hist (1)&2156&~~10&~~36.470&~152.400&~~56.930&~~~7.401\\ 
&+ cpm&2156&~~~6&~~34.960&~~71.550&~~65.880&~~~7.548\\ 
&+ hist by gender&2156&~~12&~~36.260&~143.600&~~38.590&~~~7.426\\ 
&+ 2Dhist&2156&~~18&~~36.320&~105.700&~~34.320&~~~7.405\\ 
&+ hist by WE&2156&~~11&~~35.760&~127.100&~~24.180&~~~7.470\\ \hline
(e)&base&1656&~~~5&~~32.090&~~~0&~~~0&~~~7.852\\ 
&+ hist (1)&1656&~~~9&~~35.340&~109.800&~~24.820&~~~7.824\\ 
&+ cpm&1656&~~~6&~~33.720&~~39.020&~~33.620&~~~7.873\\ 
&+ hist by gender&1656&~~11&~~35.200&~104.100&~~~7.449&~~~7.839\\ 
&+ 2Dhist&1656&~~14&~~33.560&~~27.450&~-20.210&~~~7.836\\ 
&+ hist by WE&1656&~~~9&~~34.630&~~91.160&~~~2.812&~~~7.862\\ \hline
(f)&base&~963&~~~5&~~33.940&~~~0&~~~0&~~~8.048\\ 
&+ hist (1)&~963&~~~8&~~34.810&~~38.970&~-34.450&~~~8.011\\ 
&+ cpm&~963&~~~6&~~34.420&~~~6.101&~~~1.232&~~~8.018\\ 
&+ hist by gender&~963&~~11&~~35.590&~~47.660&~-40.110&~~~7.899\\ 
&+ 2Dhist&~963&~~13&~~35.150&~~10.110&~-27.770&~~~7.954\\ 
&+ hist by WE&~963&~~~9&~~34.860&~~38.630&~-39.760&~~~7.843\\ 
\hline
\end{tabular}
\vspace{3mm}
\end{center}
\end{table}
%
The resulting estimates of the coefficient function $f(p_j)$  for the best model~(\ref{modbasic}) for each of the 6 different age combinations (a-f) in Figure~\ref{fig:est} clearly show that the effect of PA is not linear over intensity. The  $f(p_j)$ can be interpreted as a weight in the approximation to the integral of the histogram (with the first bin removed). With sedentary activity (counts 0 to 100 cpm) as a reference, the estimated weight function is positive in the range of the accelerometer profile which has a increasing contribution to fat mass and negative in the range of accelerometer counts with a decreasing contribution to fat mass. Compared to spending time in sedentary activity, time spent in activity with intensity above 1400 cpm has a decreasing effect on log fat mass. Between 200 and 800 cpm the effect of activity is increasing. 

Using the 'best' model~(\ref{modbasic}) we investigate how much variability in log fat mass the separate terms explain in Table~\ref{Tab:terms}. Overall all models~(\ref{modbasic}) have an adjusted R$^2$ between 28\% and 36\%. It is interesting to see that the ranking of effects changes with age. For the cross-sectional model at age 12 (a) $f({\tt height})$ explains most of the variability in model~\ref{modbasic}, for all other age combinations, (b) - (f) with the outcome log fat mass at ages 14 or 16, sex is the most important term and dropping sex halves the adjusted R$^2$.  This has most likely to do with physiological changes through adolescence where the percentage body fat of girls increases compared to boys. Except for model (f) PA is the second most important predictor in terms of explaining variability. 
%
\begin{table}[!h]
\begin{center}
\caption{The effect of dropping each term in turn from model~\ref{modbasic} in models for the six age combinations (a) - (f) as described in Table~\ref{Tab:models} are fitted. edf are the effective degrees of freedom of the model, R.sq is the adjusted R$^2$. Parameters, adj. R$^2$, AIC and BIC are estimated from the training data with sample size N$_T$. For AIC and BIC the difference to the 'base' model is given. The root mean squared prediction error (RMSPE) is estimated in the validation data.}
\label{Tab:terms}
\small 
\begin{tabular}{|l|l|c|c|c|c|c|c|} \hline
\multicolumn{1}{|l|}{age comb.}&\multicolumn{1}{|l|}{model}&\multicolumn{1}{|c|}{N$_T$}&\multicolumn{1}{c|}{edf}&\multicolumn{1}{c|}{adj. R$^2$}&\multicolumn{1}{c|}{$\Delta$ AIC}&\multicolumn{1}{c|}{$\Delta$ BIC}&\multicolumn{1}{c|}{RMSPE}\\ \hline
(a)&model 1&2918.0000&~~13.0000&~~31.2700&~~~0.0000&~-84.5600&~~~5.6740\\ 
&-sex&2918.0000&~~12.0000&~~29.8600&~~26.0800&~-53.8400&~~~5.7250\\ 
&-m.obese&2918.0000&~~12.0000&~~29.0600&~-91.7300&-170.8000&~~~5.7470\\ 
&-weartime&2918.0000&~~12.0000&~~31.1800&~~81.4600&~~~2.7690&~~~5.6720\\ 
&-$\sum_j f(p_j) z_{i}(p_j)$ &2918.0000&~~~8.0000&~~25.0800&-246.7000&-301.9000&~~~5.8950\\ 
&-$f_2({\tt height}_i)$&2918.0000&~~~9.0000&~~15.1000&-527.9000&-587.5000&~~~6.1560\\ \hline
(b)&model 1&1736.0000&~~12.0000&~~28.0500&~~~0.0000&~-68.9100&~~~6.4420\\ 
&-sex&1736.0000&~~12.0000&~~16.8100&-183.0000&-251.7000&~~~6.9110\\ 
&-m.obese&1736.0000&~~11.0000&~~25.7400&~-53.8100&-116.9000&~~~6.5230\\ 
&-weartime&1736.0000&~~11.0000&~~27.9500&~~67.3600&~~~3.7820&~~~6.4540\\ 
&-$\sum_j f(p_j) z_{i}(p_j)$ &1736.0000&~~~6.0000&~~23.3300&-105.2000&-145.9000&~~~6.6250\\ 
&-$f_2({\tt height}_i)$&1736.0000&~~~9.0000&~~24.6200&~~-9.4460&~-64.4100&~~~6.6110\\ \hline
(c)&model 1&2156.0000&~~10.0000&~~36.4700&~~~0.0000&~-61.4500&~~~7.4010\\ 
&-sex&2156.0000&~~13.0000&~~16.4700&-531.3000&-608.5000&~~~8.1270\\ 
&-m.obese&2156.0000&~~~9.0000&~~34.5200&~-64.0000&-119.1000&~~~7.5850\\ 
&-weartime&2156.0000&~~~9.0000&~~36.3600&~~58.7300&~~~3.2220&~~~7.4010\\ 
&-$\sum_j f(p_j) z_{i}(p_j)$ &2156.0000&~~~5.0000&~~32.7300&-118.4000&-152.4000&~~~7.5640\\ 
&-$f_2({\tt height}_i)$&2156.0000&~~~9.0000&~~33.5600&~-33.9000&~-88.4800&~~~7.5340\\\hline 
(d)&model 1&1630.0000&~~10.0000&~~27.9000&~~~0.0000&~-59.2500&~~~6.9560\\ 
&-sex&1630.0000&~~12.0000&~~15.4500&-202.7000&-273.4000&~~~7.2830\\ 
&-m.obese&1630.0000&~~~9.0000&~~25.3700&~-55.2000&-108.0000&~~~7.0950\\ 
&-weartime&1630.0000&~~~9.0000&~~27.6600&~~54.7300&~~~0.9018&~~~6.9220\\ 
&-$\sum_j f(p_j) z_{i}(p_j)$ &1630.0000&~~~7.0000&~~24.1800&~-78.9700&-121.5000&~~~7.0550\\ 
&-$f_2({\tt height}_i)$&1630.0000&~~~7.0000&~~25.2300&~~~2.8650&~-40.1300&~~~7.1070\\ \hline
(e)&model 1&1656.0000&~~~9.0000&~~35.3400&~~~0.0000&~-52.5100&~~~7.8240\\ 
&-sex&1656.0000&~~13.0000&~~15.1900&-400.6000&-474.9000&~~~8.6350\\ 
&-m.obese&1656.0000&~~~8.0000&~~33.1000&~-55.3200&-102.2000&~~~8.0030\\ 
&-weartime&1656.0000&~~~7.0000&~~35.1700&~~49.5300&~~~3.7420&~~~7.8060\\ 
&-$\sum_j f(p_j) z_{i}(p_j)$ &1656.0000&~~~5.0000&~~32.0900&~-77.3400&-109.8000&~~~7.8520\\ 
&-$f_2({\tt height}_i)$&1656.0000&~~~8.0000&~~32.8200&~~-9.7770&~-57.6700&~~~7.9590\\ \hline
(f)&model 1&~963.0000&~~~8.0000&~~34.8100&~~~0.0000&~-44.2000&~~~8.0110\\ 
&-sex&~963.0000&~~~9.0000&~~13.4800&-229.4000&-278.9000&~~~8.2790\\ 
&-m.obese&~963.0000&~~~9.0000&~~33.0300&~-26.9500&~-76.3600&~~~7.9400\\ 
&-weartime&~963.0000&~~~8.0000&~~34.9700&~~46.9600&~~~4.8210&~~~7.9860\\ 
&-$\sum_j f(p_j) z_{i}(p_j)$ &~963.0000&~~~5.0000&~~33.9400&~~-9.7500&~-38.9700&~~~8.0480\\ 
&-$f_2({\tt height}_i)$&~963.0000&~~~9.0000&~~33.6500&~~26.3400&~-22.4800&~~~7.9080\\ 
\hline
\end{tabular}
\vspace{3mm}
\end{center}
\end{table}

\subsection{Predicting fat mass with a redistributed PA}
\begin{figure}[!h]\centering
\includegraphics[scale=0.4,angle=-90]{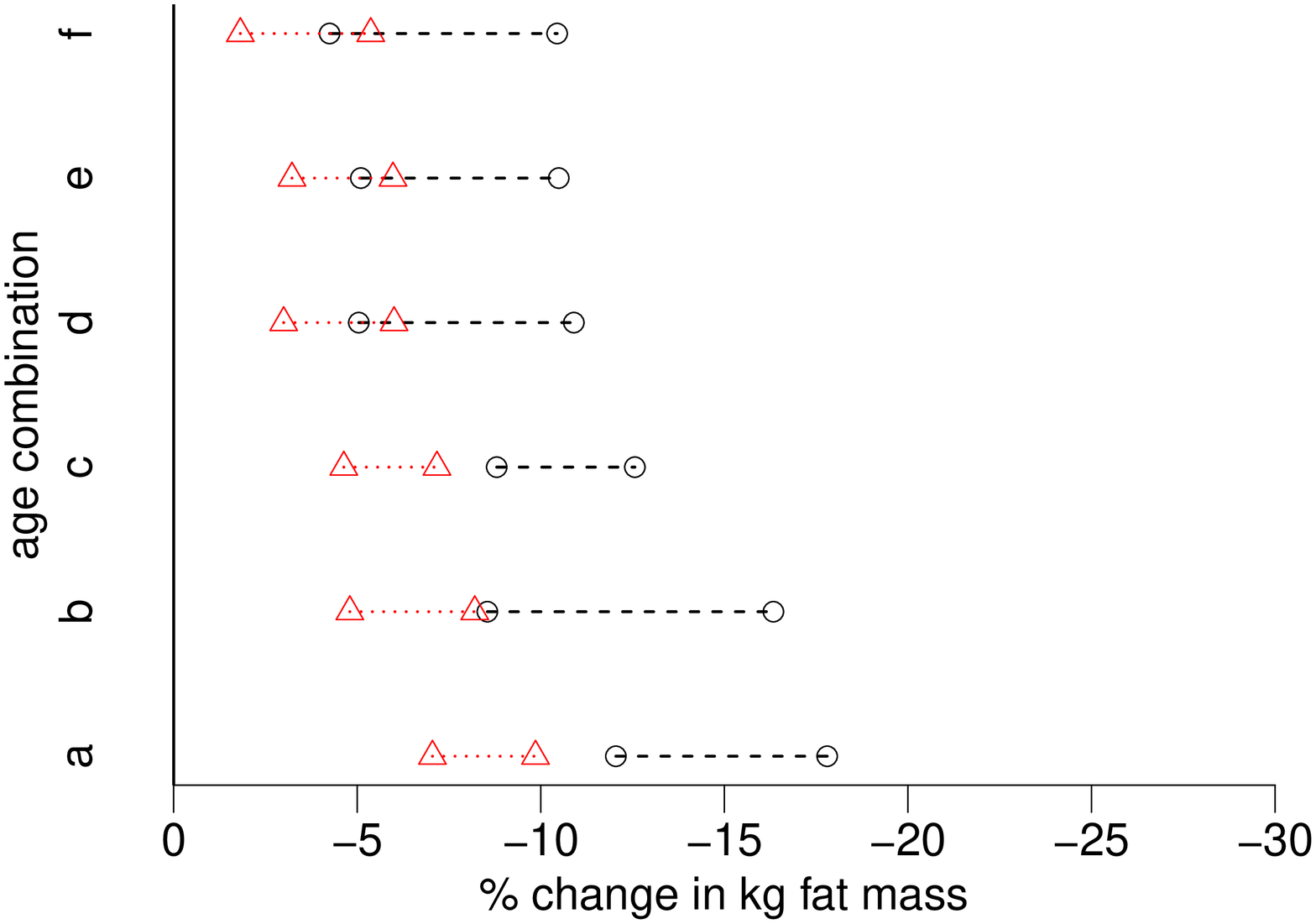}
\includegraphics[scale=0.4,angle=-90]{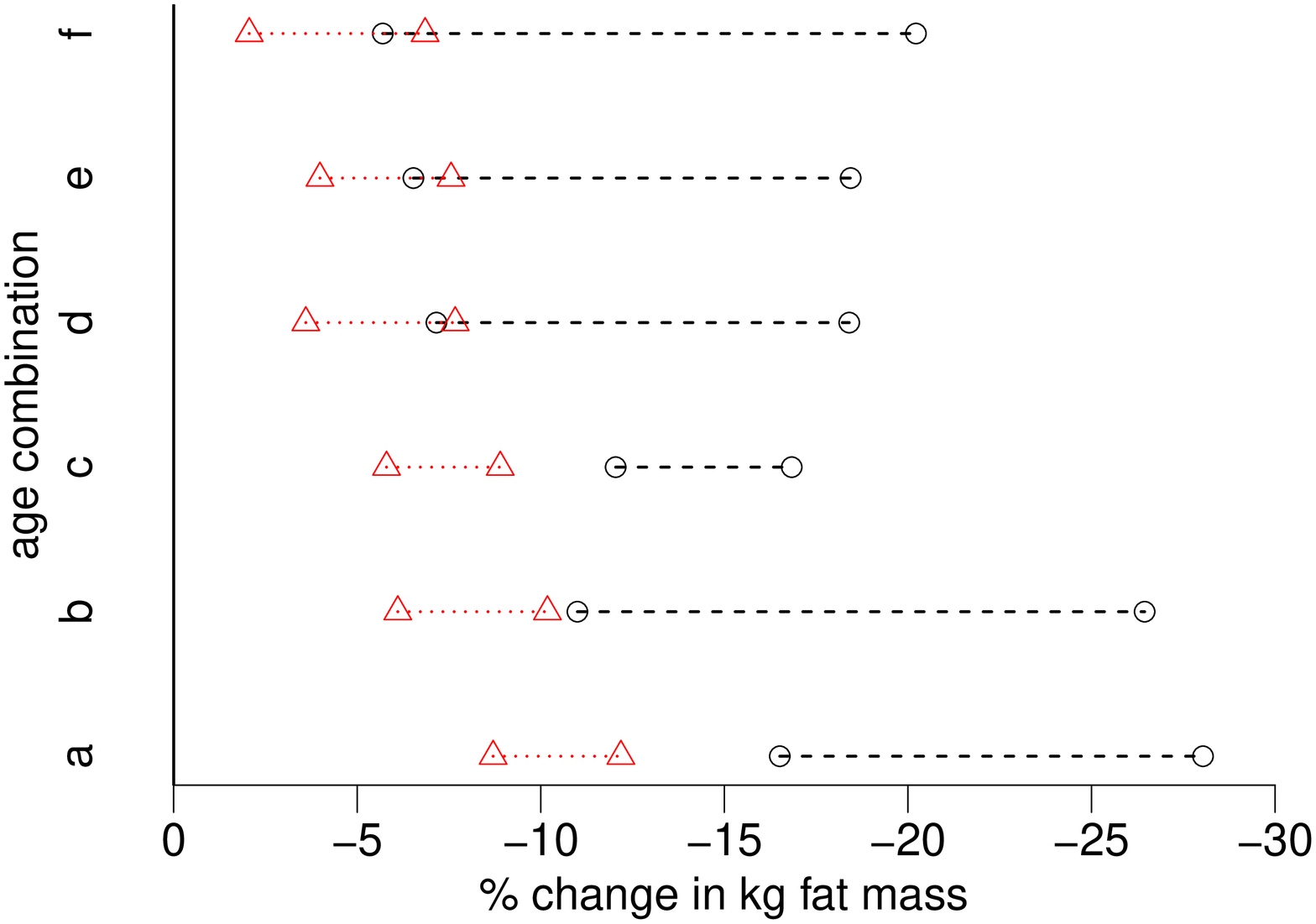}
\caption{The 95\% credible intervals for the percentage change in kg fat mass predicted for the best model~(\ref{modbasic}) are shown with a black dashed line for each of the 6 different age combinations (a-f). The red dotted line shows the 95\% credible intervals for the percentage change in kg fat mass predicted from model~\ref{modlin}. Top: Scenario 1, predicted for a redistribution of 15 minutes activity evenly into activities between moderate to vigorous. Bottom: Scenario 2, predicted for a redistribution of 15 minutes sedentary (0 - 100 cpm) activity evenly into vigorous activities.
\label{CIplot}}
\end{figure}

 We predict the percentage change in fat mass when taking away 15 minutes of sedentary activity  (0 - 100 cpm) per day and redistributing this time evenly at intensities considered as moderate to vigorous (above 3600 cpm, scenario 1) and at intensities considered as vigorous (scenario 2). A comparison of confidence intervals from model~(\ref{modbasic}) and model~(\ref{modlin}) shown in Figure~\ref{CIplot} shows that model~(\ref{modbasic}) always gives higher predictions of the percentage reduction in fat mass than model~(\ref{modbasic}). Scenario 2 achieves a higher reduction in fat mass than scenario 1. A redistributed PA at age 12 according to scenario 2 achieves a predicted mean reduction of 14.3\% of fat mass at age 16 (combination c) with a 95\% credible interval between  (12 - 17 \%).

\section{Discussion}
\label{sec:discuss}
Our approach is useful for exploring the whole range of PA when modelling fat mass. Our method allows us to consider the entire distribution of accelerometer counts in the model without any of the collinearity problems introduced when summaries of specific activity levels are used simultaneously in the model.

We have shown that the proposed methodology can be used as a diagnostic tool to evaluate whether specific single summary statistics for PA are appropriate as predictors for a specific health outcome. For instance, using the summary total activity, defined as mean daily cpm, is equivalent to assuming a linear effect over the activity range. Using MVPA assumes zero effect of activity for an intensity below moderate and a constant effect of activity at and above the assumed cut-point for moderate intensity.  Figure~\ref{fig:est} clearly shows that the effect of PA is not linear and not constant over the activity range. Our comparison of the RMSPE estimated by validation confirms, that the proposed models yield a lower prediction error than the model with mean daily cpm.  
See also \cite{AugMatCoop12} with similar findings based on a subset of the ALSPAC data. 

We have also demonstrated how to answer specific questions about the effect of PA intensity on a particular outcome. We can answer the 4 questions we posed in the introduction: 
There is evidence that the effect of PA is not linear over the intensity range. Specifically there are significant associations between increased fat mass and PA at different lower light intensity levels (300 - 1000 cpm)  as the 95\% credible intervals of $f(z_j)$ did not include zero at these intensity levels.  This intensity  corresponds to light activities such as slow walking (e.g., to shops) or general hanging out with friends, which again may be associated with certain (eating) habits.
In the prospective models (b, c, e) and the cross-sectional model at age 16 (f) we found that increased PA intensity of 2400 cpm and higher is  associated with decreased fat mass (Figure~\ref{fig:est}). 
For the two cross-sectional models at age 12 and 14 (a, d), the confidence intervals are wider, but it appears that the decreasing effect on log fat mass and PA starts at a higher intensity of around 4000 cpm. 
It is important to note that these results do not imply causality as our data are observational. In fact causal effects in both directions between PA and fat mass are possible. In particular adiposity may lead children to be less active.

 For most age combinations there is little evidence to suggest that the effect of PA intensity varies by gender (model~\ref{modsex}), whether it happens on weekdays or on weekends (model~\ref{modWE}) or whether it varies by time of the day (model~\ref{mod2D}). For the age combinations (b) the RMSPE of model~\ref{modbasic}, \ref{modsex} and \ref{mod2D} are very close and for the cross-sectional model at age 16 (f) model~\ref{modWE} has the lowest RMSPE.
This is in contrast to \cite{Ness07} where an interaction between the effects of gender and MVPA was reported from an analysis of the ALSPAC data using a cross-sectional model, with both obesity and PA at age 12. These different results can be explained by the fact that quite different models are fitted. Our approach uses the histogram of PA counts as a predictor. Figure~\ref{fig:sumfun} showed that the mean histograms are quite different between boys and girls and our approach takes the shape of the PA intensity distribution into account. In \cite{Ness07} only MVPA is used as a predictor and the intensity distribution is not taken into account. The difference in estimates for MVPA in~\cite{Ness07} for boys and girls may be due to the fact that the PA intensity distribution is different for boys and girls and that the distribution of intensity has an effect on obesity.  

The model results are robust to changes in the number of bins. We selected the number of bins which minimised the BIC criterion. When the number of bins was increased further model results became unstable. The histograms of PA are highly skewed with a spike at the first bin, but since we do not use the first bin this is not a problem. We obtained very similar results from an analysis were we transform the scale of the counts by raising them to the power of 0.35. On this scale, an even spacing of bin widths results in a more uniform distribution, which is also more conducive to graphical display. 

\cite{Masse05} report on a sensitivity analysis comparing different processing protocols and conclude that the decision rules used to process accelerometer data have an impact on outcome variables and sample size. Here changing block size from 10 zeros to 60 zeros to assess non-wear time makes little difference in the estimates. The analysis with 60 block zeros produces more or less the same graph as in Figure~\ref{fig:est}. 

It is further work to incorporate this functional predictor approach into a longitudinal model. It would also be of interest to investigate whether the functional summaries can be used to classify individuals into groups with a similar activity pattern. 
Since there is evidence from many physiological studies that certain frequency patterns of PA at certain levels are beneficial, it would be of interest to confirm these findings with our modelling approach. Our method has potential to be useful for exploring such patterns. Applying our approach to other outcomes and/or high frequency count data would also be of interest. The strength and shape of the relationship between PA and health outcome may differ by outcome.  In particular we might expect to see different effects for the health outcomes blood pressure \citep{Leary08} or depression \citep{Wiles11}.

Instead of the histogram we could have used the kernel density estimate as a summary function.
Both the histogram and the kernel density estimate summarise the distribution, with the kernel density estimate being smoother but also requiring user input as the degree of smoothness has to be decided on by choosing a window width.

In ALSPAC PA levels were measured for a relatively short time, a minimum of three days, and we assume that the measurements reflect the individuals typical activity pattern. There are some exceptions, as water based activities cannot be measured by the accelerometers used and some activities such as cycling are not well measured. Given the large sample size, we assume that these exceptions are negligible.  The reliability and representativeness of accelerometer measurements is also supported by results from a repeated measures analysis \citep{MattLeaNe07} with  individuals having repeats of wearing accelerometers. This study showed that the intra class coefficient (ICC) was moderately high for certain scalar summaries, e.g. the ICC for total activity was 0.54 and for MVPA it was 0.45. 

Although typical densities at high intensity levels are low, the time spent at the most intense levels of activity is very important. The estimates of $f(x_j)$ are negative and have overall the highest absolute values in that range, implying that spending a certain number of minutes at the most intensive range of activity is associated with the highest reduction of fat mass.  This has implications regarding the effectiveness of public health interventions. Figure~\ref{CIplot} shows that the potential benefit of redistributing time from sedentary to moderate to vigorous activity is substantial (scenario 1) and a redistribution of time from sedentary to vigorous activity (scenario 2) has an even a higher impact. An estimated mean reduction of 14.3\% in fat mass  at age 16 using model~(\ref{modbasic}) with a credible interval 12 - 17\% when 15 minutes of time per day spent in sedentary activity at age 12 is redistributed to vigorous activity is a modest although clinically relevant effect. Redistributing say, 60 minutes, according to scenario 2 would be associated with a more substantial fat reduction.   These predictions also show how our model can be used to investigate potential changes regarding PA. In summary, our approach allows to investigate the effect of PA intensity on a particular outcome in detail. It also allows to investigate potential effects of a PA intervention and to optimise the impact of an intervention. This has implications regarding the effectiveness of public health interventions and PA intervention studies. 

We expect that our approach will be useful for other data of similar nature to the accelerometer time series, i.e. high dimensional data produced by monitoring individuals, arising in many medical and epidemiological areas. For example our approach could be applied to investigate the effect of daily pollution or temperature data on certain health outcomes.

\section*{Acknowledgements}
We are extremely grateful to all the families who took part in this study, the midwives for their help in recruiting them, and the whole ALSPAC team, which includes interviewers, computer and laboratory technicians, clerical workers, research scientists, volunteers, managers, receptionists and nurses. The UK Medical Research Council, the Wellcome Trust (Grant ref: 092731) and the University of Bristol provide core support for ALSPAC. This publication is the work of the authors and Nicole Augustin will serve as guarantor for the contents of this paper. Part of this research was specifically funded by the NIHR methods opportunity funding scheme 2009.
We thank Ashley Cooper for his comments regarding the pre-processing of accelerometer counts. Many thanks to Alexandra Griffiths for helpful discussions regarding data processing and modelling.

Ethical approval for the study was obtained from the ALSPAC Ethics and Law Committee and the Local Research Ethics Committees.
\vspace*{-8pt}






\end{document}